\documentclass[10pt]{ismd08}
\usepackage{graphicx}
\usepackage{cite,./mcite}
\def\bibtex{{\sc BiBTeX}}
\begin{document}
\title{Squeezed correlations among particle-antiparticle pairs}
\author{Sandra S. Padula$^1$\protect\footnote{\ \ speaker}, Danuce M. Dudek$^1$,
 Ot\'avio Socolowski Jr.$^2$}
\institute{$^1$IFT-UNESP, Rua Pamplona, 145, 01405-900 S\~ao Paulo, SP, Brazil,\\ 
$^2$IMEF - FURG - Caixa Postal 474, 9620-900, Rio Grande, RS, Brazil}
\maketitle
\begin{abstract}
The hadronic correlation among particle-anti\-particle pairs was highlighted in the late 1990's, culminating with 
the demonstration 
that it should exist if the masses of the hadrons were modified in the hot and dense  
medium formed in high energy heavy ion collisions. They were called Back-to-Back Correlations (BBC) of particle-antiparticle pairs, also known as squeezed correlations. However, even though they are well-established theoretically, such hadronic correlations have not yet been  experimentally discovered.  Expecting to compel the experimentalists to search for this effect, we suggest here a clear way to look for the BBC signal, by constructing the squeezed correlation function of $\phi\phi$ and $K^+ K^-$ pairs at RHIC energies, plotted in terms of the average momentum of the pair,  $\bf {K_{12}}\!\!=\!\!\frac{1}{2} {\bf (k_1 + k_2)} $, inspired by procedures adopted in Hanbury-Brown \& Twiss (HBT) correlations. 

\end{abstract}

\section{Basic Formalism}

Back-to-Back Correlations (BBC) of particle-antiparticle pairs, also called hadronic squeezed correlations, 
were predicted to exist if their masses were modified in the hot and dense medium formed in high 
energy heavy ion collisions. The formalism corresponding to the bosonic case was developed in Ref.\cite{acg99}. 
Shortly after that, similar correlations were shown to exist among fermion-antifermion pairs with 
in-medium modified masses\cite{pchkp01}, and they were treated by an analogous formalisms. 
However, in contrast to what is observed in quantum statistical 
correlations of identical hadrons  (the HBT effect), where bosons with similar momenta have positive 
correlations, while fermions with similar momenta are anti-correlated, the fermionic (fBBC) and the 
bosonic (bBBC)  Back-to-Back Correlations are both positive correlations with unlimited intensity. 
The similarities of the fBBC and the bBBC 
curves were illustrated in Fig. 1 of Ref.\cite{pchkp01}, where squeezed correlations of two 
$\phi$-mesons and of $\bar{p}p$ were chosen as illustration. 
In what follows, we will focus our discussion in the bosonic case, illustrating the effect by considering 
$\phi \phi$ pairs, and also introducing some results on $K^+ K^-$ pairs. 

In the case of $\phi$-mesons (which are their own antiparticles)  with in-medium modified masses,  
the joint probability for observing two such particles, i.e., the two-particle distribution, is written as 
$N_2(\mathbf k_1,\mathbf k_2)\! =\!\omega_{\mathbf k_1} \omega_{\mathbf k_2} \, \Bigl[\langle
a^\dagger_{\mathbf k_1} a_{\mathbf k_1}\rangle \langle a^\dagger_{\mathbf k_2}
a_{\mathbf k_2} \rangle + \langle a^\dagger_{\mathbf k_1} a_{\mathbf k_2}\rangle
\langle a^\dagger_{\mathbf k_2} a_{\mathbf k_1} \rangle + \langle
a^\dagger_{\mathbf k_1} a^\dagger_{\mathbf k_2} \rangle \langle a_{\mathbf k_2}
a_{\mathbf k_1} \rangle\Bigr]$,  after applying a generalization of Wick's theorem for locally
equilibrated systems. 
The first term corresponds to the product of the spectra of the two $\phi$'s, 
$N_1(\mathbf k_i)\!=\!\omega_{\mathbf k_i} \frac{d^3N}{d\mathbf k_i} \!=\!
\omega_{\mathbf k_i}\,
\langle a^\dagger_{\mathbf k_i} a_{\mathbf k_i} \rangle $, 
being $a^\dagger_\mathbf k$ and $a_\mathbf k$ the free-particle creation and
annihilation operators of the scalar quanta, and 
$\langle ... \rangle $ means thermal averages. 
The second term contains the identical particle contribution which, together with the first term, 
gives rise to the femtoscopic (or Hanbury-Brown \& Twiss) effect, and is represented by the 
square modulus of the so-called chaotic amplitude,
$ G_c({\mathbf k_1},{\mathbf k_2}) = \sqrt{\omega_{\mathbf k_1} \omega_{\mathbf k_2}} \; \langle
a^\dagger_{\mathbf k_1} a_{\mathbf k_2} \rangle$. 
The third term, when written as the square modulus of the squeezed amplitude, 
$G_s({\mathbf k_1},{\mathbf k_2}) = \sqrt{\omega_{\mathbf k_1} \omega_{\mathbf k_2} } \; \langle
a_{\mathbf k_1} a_{\mathbf k_2} \rangle$, 
is identically zero in the absence of in-medium mass-shift. However, if the masses of the particles are modified,  
it gives rise to the squeezed correlation function, together with the first term. %
In summary, in terms of these amplitudes, the $\phi \phi$ correlation function can be written as
\begin{equation}
 C_2({\mathbf k_1},{\mathbf k_2}) =1 + 
\frac{|G_c({\mathbf k_1},{\mathbf k_2})|^2}{G_c({\mathbf k_1},{\mathbf k_1}) G_c({\mathbf k_2},{\mathbf k_2})}
+ \frac{|G_s({\mathbf k_1},{\mathbf k_2})|^2}{G_c({\mathbf k_1},{\mathbf k_1}) G_c({\mathbf k_2},{\mathbf k_2})}, \label{fullcorr}
\end{equation}
the first two terms corresponding to the identical particle (HBT) correlation, whereas the first and 
the last terms represent the correlation function between the particle and its antiparticle, i.e., the squeezed part. 
In the case of charged mesons, as in the $K^+ K^-$, only the first and the 
last terms contribute to the squeezed correlation, if their masses change. 

In the definition of the amplitudes $ G_c({\mathbf k_i},{\mathbf k_j}) $ and $ G_s({\mathbf k_i},{\mathbf k_j}) $, the annihilation (creation) operator of the asymptotic, observed bosons with momentum 
$k^\mu\!=\!(\omega_k,{\bf k})$, i.e., $a$ ($a^\dagger$), is related to the in-medium annihilation (creation) 
operator $b$ ($b^\dagger$), corresponding to thermalized quasi-particles, by the Bogoliubov-Valatin 
transformation,  
\begin{equation}
a_k=c_k b_k + s^*_{-k} b^\dagger_{-k} \;\; ; \;\;a^\dagger_k=c^*_k
b^\dagger_k + s_{-k} b_{-k} \;\; ; \;\;  f_{i,j}(x)=
\frac{1}{2}\log\left[\frac{K^{\mu}_{i,j}(x)\, u_\mu
(x)} {K^{*\nu}_{i,j}(x) \, u_\nu(x)}\right]. \label{bogovalan}
\end{equation}
In Eq. (\ref{bogovalan}), $c_k \equiv \cosh(f_k)$, $s_k \equiv \sinh(f_k)$. The argument,  $f_k$, is 
called  {\sl squeezing parameter}, since the transformation in Eq. (\ref{bogovalan}) is equivalent to a 
squeezing operation. The in-medium modified mass, $m_*$, is related to the asymptotic mass, 
$m$, by $m_*^2(|{\bf k}|)=m^2-\delta M^2(|{\bf k}|)$. Although in the general case $m_*$ could be momentum-dependent,  
it is here assumed to be a constant mass-shift. For a hydrodynamical ensemble, both the chaotic and the squeezed
amplitudes, $G_c$ and $G_s$, respectively, can be written in a special form derived 
in \cite{sm}, and developed in \cite{acg99,phkpc05}. 
\section{\bf Results} 
\begin{figure}[htb]
\begin{center}
 \includegraphics[height=.28\textheight]{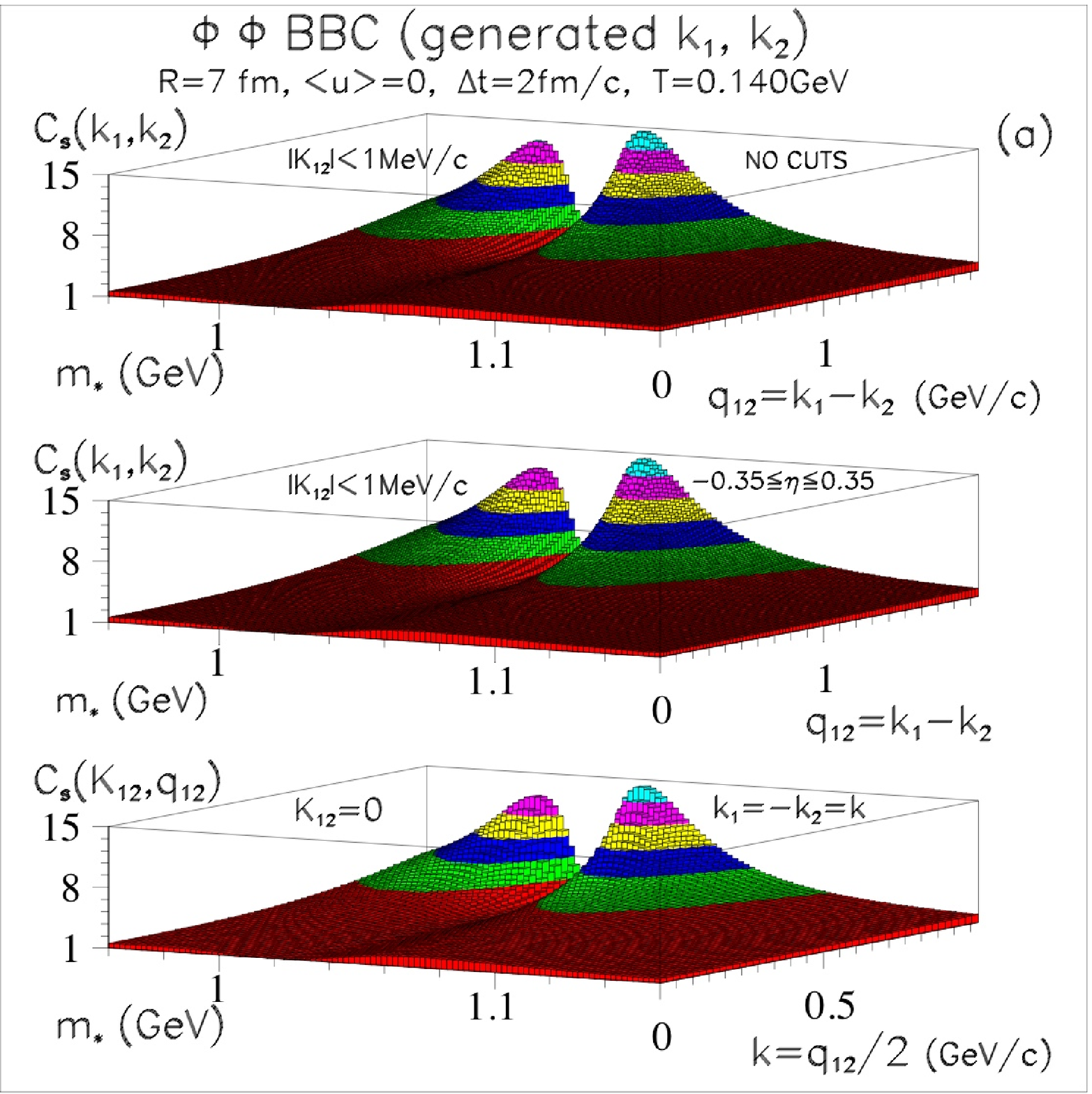}
  \includegraphics[height=.28\textheight]{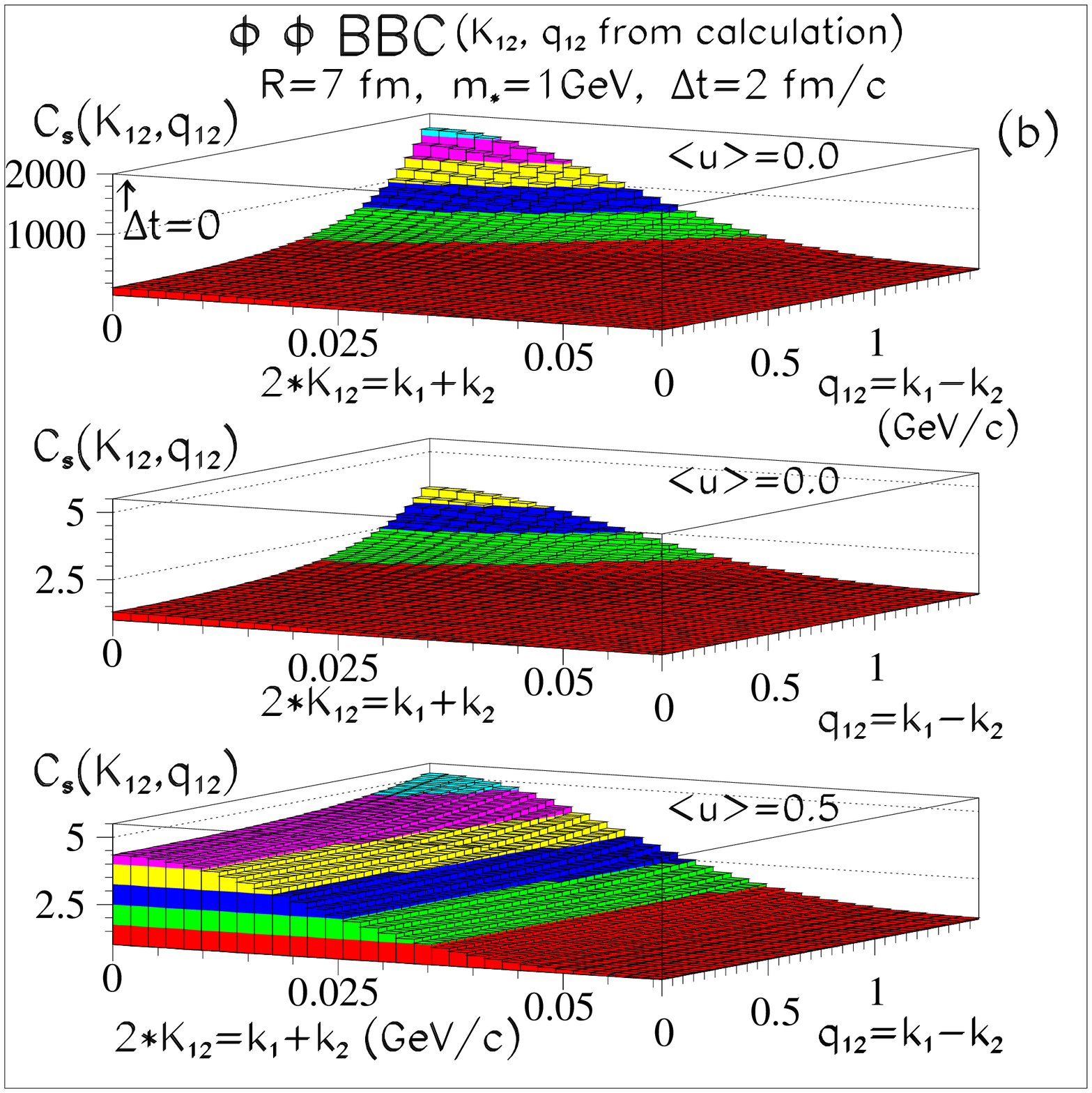}
 \caption{Part (a) shows the squeezed-pair correlation as a function of the in-medium mass, $m_*$, and of the back-to-back momentum of each particle, for a static medium ($\langle u \rangle\!=\!0$)).  In  (b) the effects of finite emission time ($\Delta t=2$fm/c) and of radial flow ($\langle u \rangle\!=\!0.5$) are shown, for fixed $m_*=1$GeV .}\label{corrphi}
\end{center}
\end{figure}

The formulation for both bosons and fermions was initially derived for a static, infinite medium \cite{acg99,pchkp01}. More  recently, it was shown\cite{phkpc05}  in the bosonic case that, even for finite-size systems expanding with moderate flow,  the squeezed correlation may survive with sizable  strength to be observed experimentally. Similar behavior is expected in the fermionic case. In that analysis, a non-relativistic treatment with flow-independent squeezing parameter was adopted for the sake of simplicity, which allowed for obtaining analytical results. The detailed discussion is in Ref. \cite{phkpc05}, where the maximum value of $C_s({\mathbf k},-{\mathbf k})$, was studied as a 
function of the modified mass, $m_*$, considering pairs with exact back-to-back momenta, 
${\mathbf k_1}\!\!=\!\!-{\mathbf k_2}\!\!=\!\!{\mathbf k}$. This type of analysis represents an analogous procedure as to studying only the intercept parameter of the HBT correlation function. This is illustrated in Fig. \ref{corrphi}(a), which shows the variation of the maximum of the squeezed correlation in the absence of flow, in three parts. The top and middle plots are results of a recent simulation, where the momenta of each particle in the pair is generated, the squeezed correlation is then estimated and the bins are filled. The bottom plot is obtained by attributing precise values to the variables, then calculating $C_s(m_*, q_{_{12}})$. This shows that the simulation is indeed reproducing the calculation, for small bin sizes. 
We can also see from Fig. \ref{corrphi}(a) that the simulation shows practically no sensitivity to the cuts introduced in the momentum generation, in order to mimic the experimental cuts in $p_T$, $\eta$, azimuthal angle, etc\cite{phenixphi}. Although this study illustrated many points of theoretical interest, it was not helpful for motivating the experimental search of the BBC's, since the modified mass is not accessible to direct measurement.

A more realistic analysis would involve combinations of the momenta of the particles, in terms of which the BBC could be searched for, even though we would have to make a more precise hypothesis concerning the mass-shift. For the sake of simplicity and for illustrating the procedure, we will assume here a constant value for $m_*$. Within the non-relativistic approach of  \cite{phkpc05}, we suggest to combine 
the particle-antiparticle momenta, $(\mathbf k_1,\mathbf k_2)$, into the pair average momentum, $\mathbf K\!=\!\frac{1}{2}( \mathbf k_1+\mathbf k_2)$, and analyze the squeezed correlation function in terms of $|\mathbf K|$, similarly to what is done in HBT interferometry. 
The maximum of the BBC effect is reached when ${\mathbf k_1}\!=\!-\!{\mathbf k_2}\!=\!{\mathbf k}$, corresponding to $|\mathbf K|\!=\!0$. Therefore, the squeezed correlation should be investigated as $C_s({\mathbf k_1},{\mathbf k_2})=C_s({\mathbf K},{\mathbf q})$, around the zero of the average momentum. For simplicity,  we analyze here the behavior of the correlation function, detailed in  \cite{phkpc05}, by attributing values to $|\mathbf K|$ and $|\mathbf q|$, as shown in Fig. \ref{corrphi}(b),  where the in-medium mass of the $\phi$'s 
was fixed to $m_*=1.0$ GeV. 
In the top and middle plots, a static system ($\langle u\rangle=0$) was considered. By comparing these two plots, we can see the dramatic r\^ole played by the finite emission times, which reduces the BBC signal by more than two orders of magnitude. This was obtained when considering an exponential emission, leading to a Lorentzian factor {\small $F(\Delta t)=[1+(\omega_1+\omega_2)^2 \Delta t ^2]^{-1}$}, 
with $\Delta t=2$ fm/c, multiplying the second and the third terms in Eq. (\ref{fullcorr}). From Fig. \ref{corrphi}(b) we also 
see that, in the absence of flow, the squeezed correlation intensity grows faster for higher values $|\mathbf q|$ than the corresponding case in the presence of flow. However, in this last one it is stronger even for smaller values of  $|\mathbf q|$, showing that the presence of flow could help to enhance the signal. 
\begin{figure}[htb]
\begin{center}
 \includegraphics[height=.28\textheight]{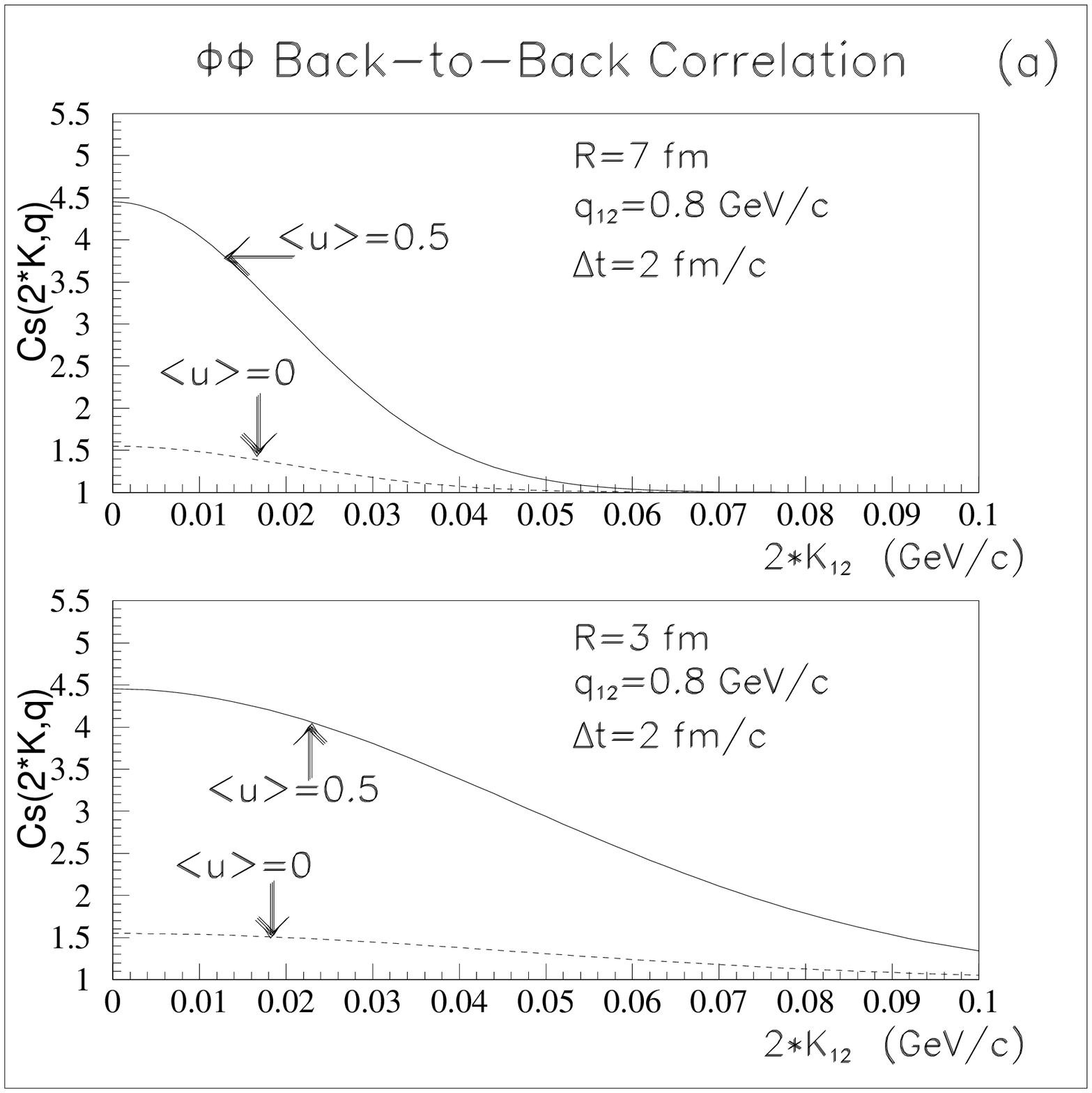}
  \includegraphics[height=.28\textheight]{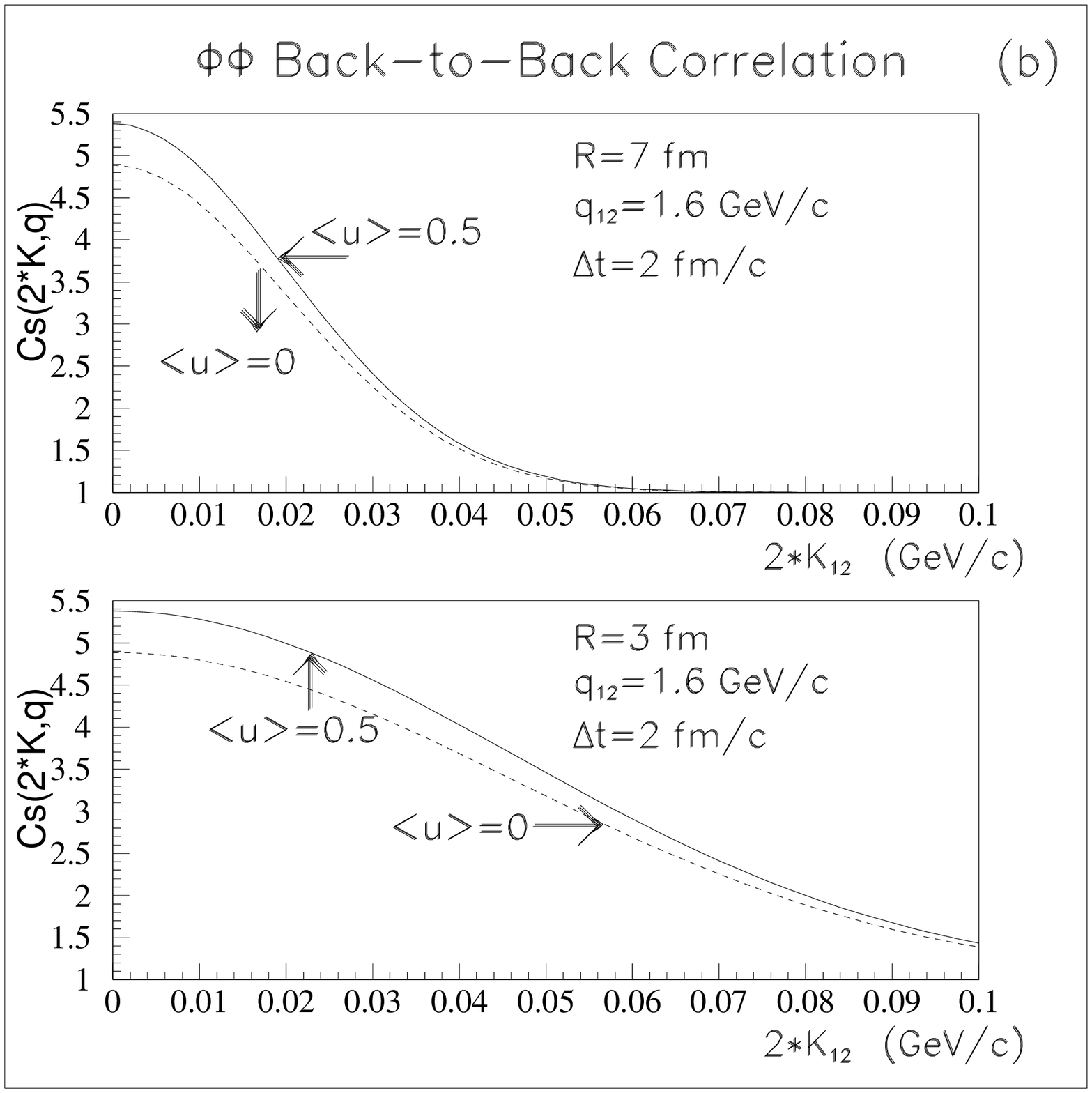}
\caption{The response of the BBC function to system sizes with $R=7$ fm (top plots) and $R=3$ fm (bottom plots) is shown, for $\Delta t = 2$ fm/c. In (a), the relative momentum was fixed to $q_{_{12}}=0.8$ GeV/c. In part (b), $q_{_{12}}=1.6$ GeV/c.}\label{r3r7}
\end{center}
\end{figure}
\begin{figure}[htb]
\begin{center}
 \includegraphics[height=.27\textheight]{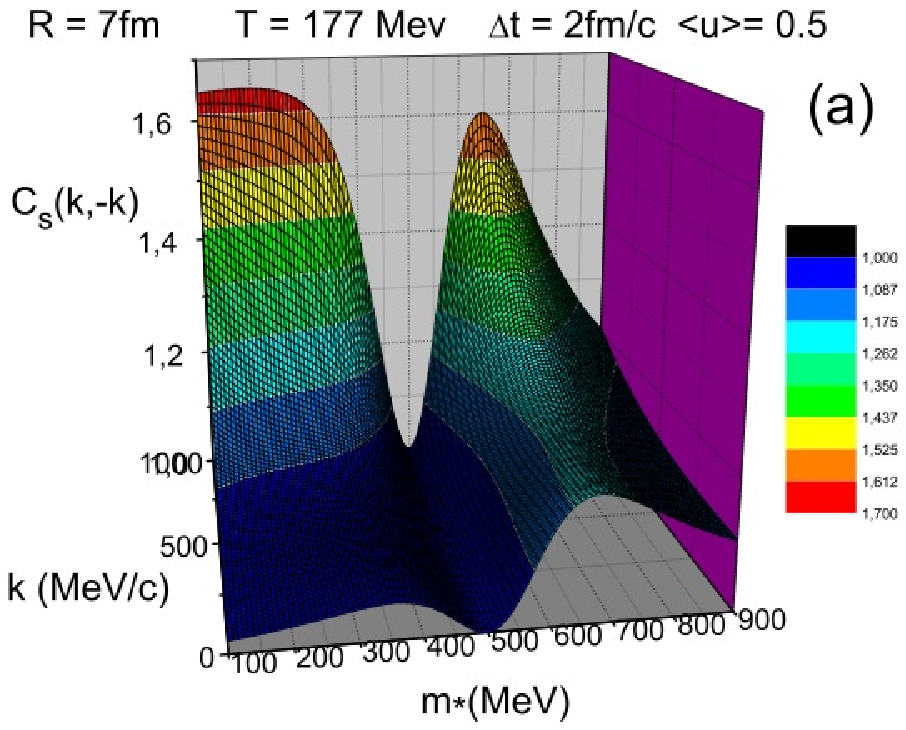}\includegraphics[height=.27\textheight]{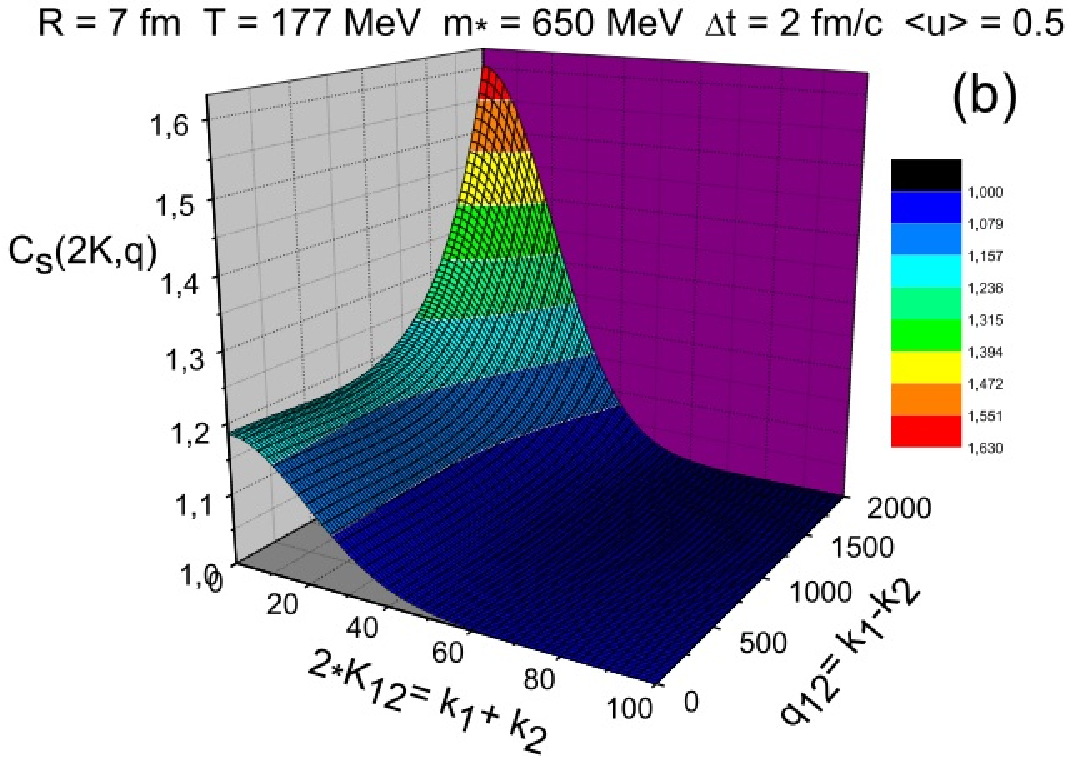}
\caption{Part (a) shows the squeezed correlation as a function of possible in-medium mass of the kaons, $m_*$, and of the momentum of each particle in the pair, $|\mathbf{k}|$. In part (b), it is plotted as a function of  $K_{_{12}}$ and $q_{_{12}}$, considering $m_*=650$ MeV, which corresponds to roughly the highest value of the correlation in part (a).}\label{kaonsqcorr}
\end{center}
\end{figure}

The sensitivity of the squeezed-pair correlation to the size of the region where the mass-change occurs is shown in Fig. \ref{r3r7}, for two values of the system radii, $R=7$ fm and $R=3$ fm, 
fixing the relative momentum of the pair to 
(a) $q_{_{12}}=0.80$ GeV/c and (b) $q_{_{12}}=1.6$ GeV/c. The plots were obtained by attributing values to $K_{_{12}}$ and  $q_{_{12}}$. 
We can see that the size of the squeezing region is reflected in the inverse width of the curves plotted as function of $2 |\mathbf K|$.  

The investigation of the squeezed correlation in terms of $2\mathbf K$ is applicable when treating non-relativistic flow. In the case of a fully relativistic study, a four-momentum variable can be constructed, as $Q_{back}=(\omega_1-\omega_2,\mathbf k_1 + \mathbf k_2)=(q^0,2\mathbf K)$, as introduced in Ref. \cite{pscn08}. Moreover, it would be preferable to redefine this variable  as $Q^2_{bbc} = -(Q_{back})^2=4(\omega_1\omega_2-K^\mu K_\mu )$, because its  non-relativistic limit recovers $Q^2_{bbc} \rightarrow (2 \mathbf K)^2$. 

The above analysis could also be applied to other particles that are not their own antiparticles. For showing it, we investigate the case of $K^+ K^-$ squeezed correlations, as illustrated in Fig. \ref{kaonsqcorr}, for an expanding system with radial flow parameter $\langle u\rangle=0.5$. In part (a),  the squeezed correlation is shown as a function of the in-medium mass, $m_*$, also varying the back-to-back momentum of particle and antiparticle. 
In part (b), the squeezed correlation is plotted as a function of ($K_{_{12}}$,$q_{_{12}}$), fixing the kaon in-medium modified mass to $m_*=650$ MeV. These plots do not come from simulation, but were obtained by attributing values to the plotting variables. 
 
\section{Conclusions}

We discussed here some of the main results on the squeezed correlations, within an a non-relativistic  approach developed earlier. We suggest an effective way to search for it in heavy ion collisions at RHIC, emphasizing the need for experimentally observe this signal. This should be done by plotting the hadronic squeezed correlations in terms of the average momentum of the pair, $(2 \mathbf K)^2$,  which is the non-relativistic limit of  the four-vector $Q^2_{bbc}=4(\omega_1\omega_2-K^\mu K_\mu )$. We showed some results that would be expected in the case of $\phi \phi$ back-to-back correlations, as well in the case of the $K^+ K^-$ pairs. We also illustrated the effects of finite system sizes, finite times and flow. We could see that finite emission times reduce the signal substantially, and that, in the presence of flow, the signal is stronger over the momentum regions in the plots, i.e., roughly for $0 \le |\mathbf K_{_{12}}| \le 60$ MeV/c and $ |\mathbf q_{_{12}}| \le 2000$ MeV/c, suggesting that flow may enhance the chances of observing the BBC signal. We also saw that the correlation function reflects the size of the region where the squeezing occurred. Finally,  we should emphasize that the absence of squeezing, i.e., if there is no in-medium mass modification, the squeezed correlation functions would be unity for all values of $2 |\mathbf K_{_{12}}|$ and $|\mathbf q_{_{12}}|$. 

\subsection{Acknowledgments}
SSP is very grateful to the Organizing Committee of the ISMD 2008, and specially to Hannes Jung, for the kind support to attend the symposium, thus making her participation possible. DMD thanks CAPES and FAPESP for the financial support.  


\providecommand{\etal}{et al.\xspace}
\providecommand{\href}[2]{#2}
\providecommand{\coll}{Coll.}
\catcode`\@=11
\def\@bibitem#1{%
\ifmc@bstsupport
  \mc@iftail{#1}%
    {;\newline\ignorespaces}%
    {\ifmc@first\else.\fi\orig@bibitem{#1}}
  \mc@firstfalse
\else
  \mc@iftail{#1}%
    {\ignorespaces}%
    {\orig@bibitem{#1}}%
\fi}%
\catcode`\@=12
\begin{thebibliography}{99}
\begin{footnotesize}
\bibliographystyle{ismd08} 
{\raggedright
\bibitem{acg99} M. Asakawa, T. Cs\"org\H o and M. Gyulassy,  Phys. Rev. Lett.  {\bf 83}, 4013 (1999).
\bibitem{pchkp01}  P. K. Panda, T. Cs\"org\H o, Y. Hama, G. Krein and 
Sandra S. Padula, Phys. Lett. {\bf B512}, 49 (2001).
\bibitem{sm} A. Makhlin and Yu. Sinyukov, Sov. J. Nucl. Phys. {\bf 46}, 354 (1987);  Yu. Sinyukov, Nucl. Phys. {\bf A566}, 589c (1994). 
\bibitem{phkpc05} Sandra S. Padula, Y. Hama, G. Krein,  P. K. 
Panda, and T. Cs\"org\H{o}, Phys. Rev. {\bf C73}, 044906 (2006).
\bibitem{phenixphi} S. S. Adler et al., PHENIX Collaboration,  Phys. Rev. {\bf C72}, 014903 (2005). 
\bibitem{pscn08} Sandra S. Padula, O. Socolowski Jr., T. Cs\"org\H o, and 
M. Nargy, J. Phys. {\bf G35}: Nucl. Part. Phys., 104141 (2008).
}
\end{footnotesize}
\end{thebibliography}
\end{document}